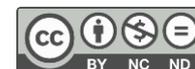

# Depth profiling the elemental composition with negative muons: Monte Carlo based tools for improved data analysis

M.Cataldo[1], A.D Hillier[2], O.Cremonesi[1], F. Grazzi[3], S. Porcinai[4], M. Clemenza[1]

[1] INFN Sezione Milano Bicocca, Milano, Italy
[2] ISIS Neutron and Muon Source, STFC Rutherford Appleton Laboratory, UK
[3] CNR-IFAC, Sesto Fiorentino, Italy
[4] Opificio delle Pietre Dure, Firenze, Italy

### Abstract

Gildings, patinas and alteration crusts are common features of many heritage artefacts, especially for metals. Their size depends on many factors, like the manufacturing method for gildings or the conservation state for alteration crusts: in some cases, it can be in the scale of the tens of microns. Such thickness would be difficult to investigate with classical non-destructive methods and would prevent getting information from the bulk of the sample. This work proposes an innovative approach for the study of multi-layered materials with the Muonic atom X-ray Emission Spectroscopy technique (μ-XES). Based on the detection of the high-energy X-rays emitted after the muon capture by the atom, this method is characterised by a remarkable penetration depth (from microns to cm). From the surface to the bulk, this technique can evaluate the variation of the elemental composition as a function of depth. The paper focuses on providing an improved interpretation of μ-XES data by coupling the analysis with the use of two Monte Carlo simulation software, GEANT4/ARBY and SRIM/TRIM. With these two software, it is possible to replicate the negative muon experiments and compare the experimental and simulated outputs to address the size of a given layer. To validate this approach, a set of standard gilded bronze and brass foils were measured at the ISIS Neutron and Muon source. From simulations, it was possible to evaluate the thickness of the superficial gold layer, with results in agreement with the preliminary SEM characterisation of the samples.

### Introduction

The evaluation of the elemental composition as a function of depth (*depth profiling*) can provide significant information to the study of a material. In Heritage Science, techniques like Rutherford Backscattering Spectrometry (RBS), X-ray Photoelectron Spectroscopy (XPS) and Time-of-Flight Secondary Ion Mass Spectrometry (ToF-SIMS) are widely used for the characterisation of thin layers at the nano or micro scale, particularly in paintings [1,2]. Although these methods are characterised by a remarkable detection limit, their probing depth is generally limited to the tens of nanometres with an upper limit of about a micron [3,4]. This range is enough for the organic layers of paintings, but not for other materials like metals. Metallic artefacts can be covered with patinas, alteration crust or gilding whose thickness can be in the scale of the tens of microns, which the above-mentioned methods cannot probe. A valid approach is provided by X-ray Fluorescence (XRF), Particle-Induced X-ray and Gamma-ray Emission (PIXE and PIGE), which can probe microns deep into a material [5,6]. However, XRF and to some extent PIXE, are mostly limited by the self-absorption of the emitted radiation [7]. Self-absorption is influenced by factors such as layer thickness, material density, and radiation energy. This means that emitted photons can be absorbed before escaping the material, especially for low-energy X-rays. This effect, for instance, could prevent the information coming from a layer deep within the sample from reaching the detector, in particular for thick samples. Moreover, they have limitations in element detection: standard XRF and PIXE only detect elements with Z > 11, while PIGE is most effective with light elements (Z < 15; for this reason, PIXE and PIGE are often used together). Therefore, to find a technique that can penetrate deep inside a material with a non-destructive approach and a multi-elemental range, it is necessary to look for a specific particle: the negative muon. When a beam of negative muons interacts with a material, muons are captured by the atoms to form the so-called "muonic atom". In this bound state, the muon rapidly cascades down to the lowest energy level in a process that, after a first part dominated by Auger electrons emission, becomes radiative with the emission of fingerprint muonic X-rays. For extensive details on the cascade process see [8,9]. The technique that exploits this phenomenon to perform elemental analysis is called Muonic atom X-ray Emission Spectroscopy (μ-XES). Developed about 40 years ago, but limited by the technologies of the time, in the last fifteen years the technique has seen an increased number of applications, with a particular interest in the characterisation of archaeological artefacts [10-16]. Conceptually, the method can be compared to XRF: the interaction of a probe with a target produces characteristic X-rays that are collected by a detection system. However, since the muon has a mass 207 times larger than the electron, when captured it experiences binding energies that are approximately 207 times larger than the one experienced by electrons. Consequently, the emitted X-rays are highly energetic (from a few keV up to the MeV scale), thus avoiding self-absorption problems and allowing the identification of low Z elements (without sample activation). For instance, the electronic Kα emission of silicon is 1.7 keV, while the corresponding muonic Kα emission is 400 keV. Moreover, low-energy muons (E < 100 GeV/c) travel a straighter path in matter and the effects of radiative energy loss are minimal, resulting in a remarkable penetration depth: in metals like copper or silver it can be up to a centimetre. The result is a technique with the unique features of being sensitive to every atom of the periodic table (with some constraints for Z < 3) and probing beneath the surface of a material to get detailed information about the bulk. At the current stage of the research, the novelty of the method offers the possibility of continuous development, both in terms of experimental setup and data processing. The aim of this work, indeed, is to provide a protocol for improved data analysis by using Monte Carlo (MC) simulation tools. MC simulations are invaluable tools for the development of several projects in the scientific community. Especially in particle physics, where the complexity and scale of experiments are always increasing, tools like GEANT4, FLUKA and MCNP are used to simulate microscopic interactions to find solutions to macroscopic problems [17-19]. From the transportation of particles to the description of the experimental setup, these software can precisely implement all the processes involved in an investigation. With the idea of providing a direct answer to the question *"How thick is this layer?"*, the work couples the classical analysis of a muonic X-ray spectrum with the use of simulation software. Here, a GEANT4-based tool called ARBY and SRIM-TRIM have been used to model a set of samples analysed at the μ-XES instrument of the ISIS Neutron and Muon Source.

## 1. Materials and methods

### 1.1 Software

GEANT4, developed by an international collaboration at CERN, is used to simulate the passage of particles through matter [18]. GEANT4 is a toolkit: users can develop their application for a specific setup or detector starting from the tools provided by the software. At the INFN section of Milano Bicocca, GEANT4 is run through a user-friendly application called GEANT4/ARBY, developed by Oliviero Cremonesi. At Bicocca, ARBY has been used and validated for many different applications as the efficiency evaluation of HPGe detectors for gamma spectroscopy or bolometric measurements for the physics of rare events [20,21]. In ARBY, simulations require a configuration file that stores all the information about the experimental setup. Parameters such as particle type, direction, and the number of events can be easily accessed through simple command-line instructions, from which the simulation is launched. The final output is processed with dedicated software to obtain a detector response function (GEANT4 version: 10.04.p03). SRIM (Stopping and Range of Ions in Matter) and TRIM (Transport of Ions in Matter), instead, is a collection of software packages for the simulation of the interaction of energetic ions with matter [22]. SRIM simulates the stopping power of ions in a material, providing information about the energy loss and the ion ranges; while TRIM simulates the trajectory of ions in a material, providing information about the ion's path (version: 2008.04). One of the main differences between the two software is the definition of the sample environment. In TRIM, the sample is described by a sequence of different layers defined by the user. ARBY, instead, provides tools for modelling the sample as it is: with a simple geometry, a copy of the real one can be reproduced. With some assumptions (the muon particle is not implemented in TRIM, so the simulation is done with a hydrogen ion with one-ninth of its mass), the two software packages can simulate a negative muon experiment. The result of a simulation is the number of stopped muons in any given layer: by assuming that the number of stopped muons is proportional to the intensity of the emitted X-rays that contribute to the full energy peak, the simulated output can be directly compared to the measured data. In ARBY, moreover, it would be possible to generate a muonic X-ray spectrum for direct comparison with a real measurement, as it is already done in XRF [23,24]. However, muon physics is not entirely implemented in GEANT4, and the generation of X-rays is not completely reliable, as reported in a recent publication [25].

### 1.2 Samples

The interest of this work is twofold: along with the validation of the software, the aim is to provide an example of the application of negative muon data analysis to the characterisation of layered samples. Therefore, the samples studied in this work are two sets of standards, in particular gilded copper alloys covered with a defined gold thickness, as reported in Table 1. Gilded artefacts are quite common in Heritage Science research and their characterisation could provide important information for restorers and curators. The application of gold to the surface of a material (*gilding*) was not only used to improve the decorative effect but also to provide resistance to corrosion and ageing. The first evidence of metal gilding dates back to the third millennium B.C., when metal objects were mechanically covered by gold foils (by folding, riveting or hammering). Then, technological development led to the introduction of a new method called *amalgam gilding* (amalgams are alloys of mercury and another metal). Developed in China in the third century B.C., the method (also known as *fire gilding* or *mercury gilding*) is based on the mixture of a well-defined ratio of mercury and gold that is slowly heated until all the gold has melted [26,27]. After cooling in cold water, the result is a paste with an overall composition of 80-90 % mercury and 10-20 % gold. This paste is applied on the surface of the metal and then heated to let mercury evaporate and leave a durable gold layer in the range of tens of microns [28]. Amalgam gilding was widely used around Europe until the 19th Century when electroplating was introduced during the Industrial Revolution. Electroplating involves the use of water-based solutions, called "galvanic baths" containing metal ions that are deposited on a surface to provide resistant metallic coatings [29]. The two sets of samples measured in this work were produced with the amalgam and the electroplating method. The firsts were realized by the *Opificio delle Pietre Dure* during a research work on the characterisation of the alloys of the south Baptistery gate door of Florence. The second, instead, was produced in collaboration with the LEA group of the chemistry department of the University of Florence.

*Table 1* – Sample description. The first sets, defined as EPA, EPB, and EPC are three electroplated foils of brass with gold deposited on a nickel substrate (Au: 98.8 ± 0.2 %). Electroplating provides very good control of the deposition process, especially for nanometre layers. In this case, where the gold layer is thicker than in usual applications, the laboratory did not guarantee complete control of the deposition. Therefore, to provide a more reliable indication of the thickness of the layer, SEM measurements were performed (at 10 different points) to provide an average thickness (sample size: 5x5x0.2 cm). The second set of samples consists of a brass (SM3) and a bronze (EM2) foil covered by an 11 ± 1 µm gold layer made with the amalgam technique (sample size: 5x4x1 cm). The samples are pieces of a larger batch, from which a fragment was taken and analysed with SEM. Therefore, as a starting information for the simulation process, it is assumed that the average thickness of the gold layer is 11 ± 1 µm, but this may not be fully representative of the two samples.

| Sample | SEM Scan | Average thickness |
|---|---|---|
| 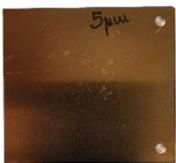 | 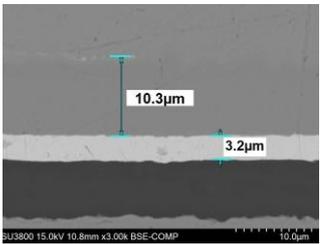 | *EP* **A** – 3.3 ± 0.2 µm<br><br>*EP* **B** – 4.6 ± 0.6 µm<br><br>*EP* **C** – 7.3 ± 0.8 µm |
| 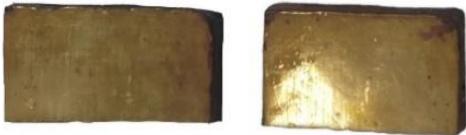 | 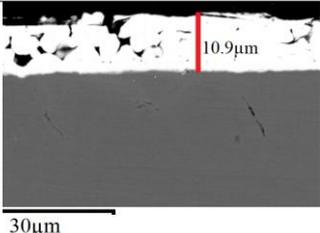 | SM3/EM2 - 11 ± 1 µm |

### 1.3 The negative muon experiment

Negative muon experiments were performed at the ISIS Neutron and Muon Source, a world leading centre for physical science located in Oxfordshire [30]. At the facility, high-energy protons are responsible for the generation of neutron and muon probes, delivered to two different target stations. Experiments were performed at Port 4 of the RIKEN-RAL facility, using the MuX setup, located in target station 1. Here, a double pulsed muon beam is delivered to the sample area, with a tunable momentum that ranges from 15 MeV/c up to 90 MeV/c. The setup for the experiment consisted of three germanium detectors placed at a defined distance from the sample position, as shown in Figure 1. Source-to-detector distance was adjusted to prevent the overloading of the detectors (i.e. event pile-up): in detail, the upstream detectors were placed at 11 and 10 cm respectively, while the downstream detector was placed at 8 cm. The sample, wrapped in aluminium, was placed in position by using an aluminium holder. Before the measurements, energy calibration was performed with radioactive sources. Finally, the set of samples was characterised by a momentum scan from 15 MeV/c up to 25 MeV/c (momentum tuning for each step varied between 0.25 and 0.5 MeV/c, but higher step change is possible) [31].

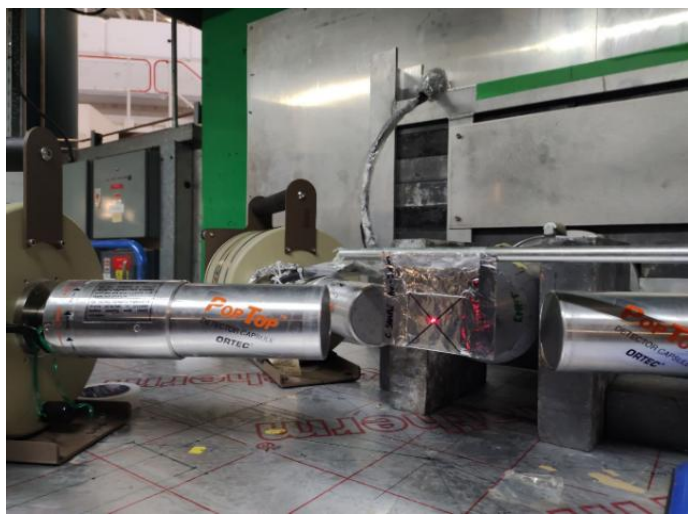

***Figure 1****- Experimental setup at port4 of the RIKEN-RAL facility: the sample, wrapped in aluminium, is placed 10 cm from the beam exit and centered with the help of a laser system. Three HPGe detectors surround the sample and are responsible for the collection of the emitted radiation.*

### 1.4 Data analysis

The acquired X-ray spectra were analysed with the Origin software (version: 2022Pro). To account for differences in measuring times, each spectrum was normalised by the total number of counts in the spectrum. Data analysis consisted of peak identification and peak fitting, done with a Gaussian function [32-34]. Most of the analysis was conducted on the output of the low-energy and high-energy upstream detectors. The first presents a high efficiency and high resolution in the low to medium energy range (0 – 500 keV), which allows a good identification of the peaks; while the second covers the high-energy range (500 keV – 8 MeV). Furthermore, by using an upstream detector, any possible self-absorption of the X-ray from the matrix is considered negligible. Each layer of the standards was identified by a given element, that corresponded to a specific peak. In detail: 102 keV for nitrogen, 134 keV for oxygen, 115 keV for copper, 130 keV for gold and 108 keV for nickel. Given that all the peaks are found in a quite narrow region of the spectrum, intensities were not corrected for the detection efficiency. A key element in a depth profile experiment is also aluminium (66 keV), used in foils as a sample holder (when not present in the measured sample). This element acts as a reference to understand where the beam is in the sample: the absence of aluminium (and nitrogen) is a clear indication the beam is probing inside the sample. It has to be mentioned that the muon beam does not have finite momentum, but its value varies within a range that is defined as spread (referred to as $\Delta p/p$ with a nominal value of 4 %). This means that the penetration of the beam is not punctual, but some muons will travel more and some less, resulting in a contribution from multiple layers that, if plotted as a function of momentum, gives the so-called "depth profile". The result of a momentum scan is shown in Figure 2, where different runs are reported. At low momentum (16 MeV/c, black spectrum), where the beam is mostly hitting the surface of the sample, aluminium and air are most responsible for the contribution to the spectrum, testified by an intense peak at 66 keV (Al), 102 keV (N) and 134 keV (O). As the penetration depth increases (with momentum), the composition starts to change until the beam is fully into the brass base layer (20 MeV/c). Finally, by plotting the value of the area under each fitted peak, a depth profile as a function of momentum is obtained, as shown in Figure 3a.

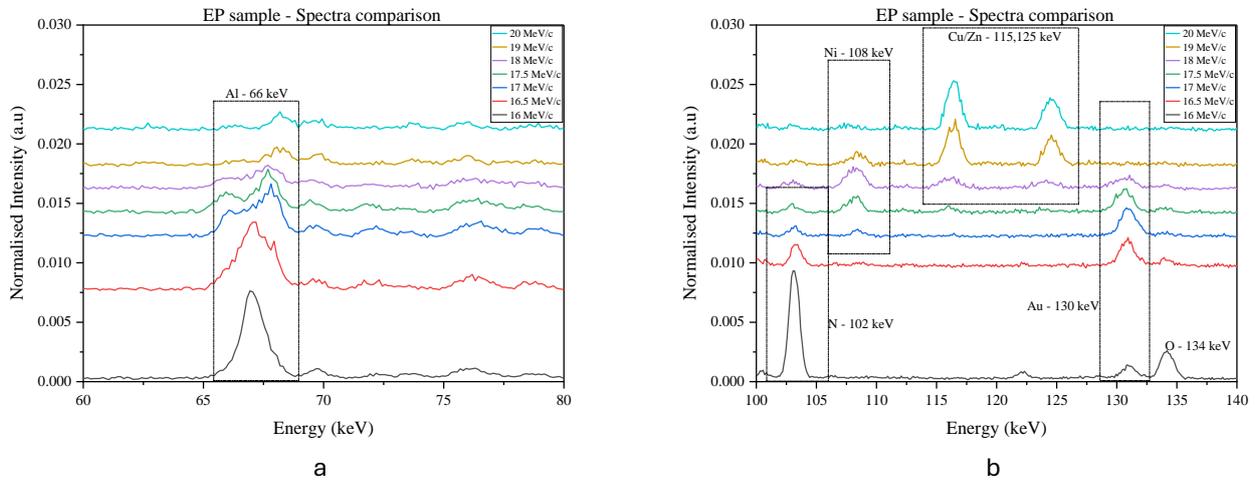

a

b

*Figure 2*- *To obtain a depth profile, runs with different momentum were performed. By changing the momentum, muons were stopped at different depths, providing specific information about the given layer. The elements used for the analysis are reported in the figures. In 2a, the aluminium peak, at 66 keV is shown (left peak in the blue and green spectra). Along with the nitrogen peak at 102 keV (2b), it gives information on where the beam is probing. When the two peaks disappear (at 17.5 MeV/c, green spectrum), the beam is fully probing inside the sample. This is testified by the following spectrum (purple), which contains information from the internal layers of the sample, like nickel (108 keV), copper (115 keV) and zinc (125 keV). Finally, at 130 keV, the gold signal, that increases with momentum and then disappears when the beam enters the nickel layer. The spectra shown in the figures are the output of the upstream low-energy detector.*

From the measured depth profile, a simulated one was reproduced with ARBY and TRIM (Fig. 3b). Simulations consisted of the interaction of a negative muon beam with a six-layer type of sample. After some preliminary tests, a good control of the simulation process was reached, and the number of events was increased ($10^6$ events) to provide more statistics. The first material to interact with muons is the Mylar window placed before the beam exit. This window is 50 μm thick and it is placed 10 cm from the sample position. Given the low density of Mylar (1.38 g/cm$^3$) and the very thin size, this layer does not contribute to the absorption of the muon beam (< 0.5 % in the momentum range used for the experiment). The second material that interacts with the beam is the 10 cm air gap that divides the Mylar window from the sample. This gap is wide enough to stop some low-energy muons, as shown in Figure 2, where the contribution from nitrogen, the main constituent of air is visible. The last four materials to interact with the beam are, in order: aluminium, gold, nickel and brass or bronze.

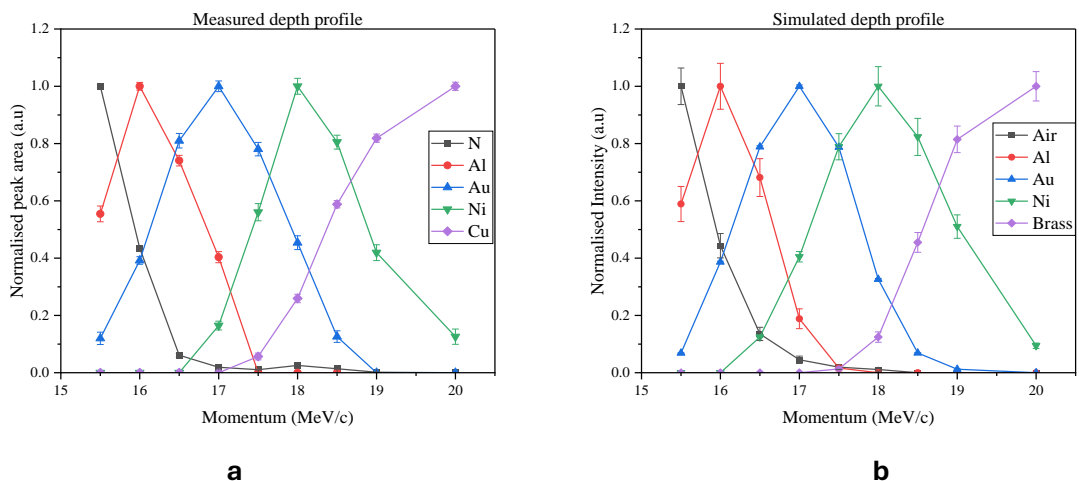

a

b

*Figure 3 – By plotting the normalised area of N, Al, Au, Ni and Cu as a function of momentum, a depth profile is obtained. This can be obtained both with the measured (a) and simulated (b) data. The comparison of the two is used to address the size of a given layer, in this case, the gilding (Au).*

## 2. Results
### 2.1 Electroplated samples

The assessment of the thickness of the gold layer consisted of a comparison between the measured and the simulated profile. For each sample, simulations with different gold layer sizes were performed, in a range of ±0.5 μm from the SEM average value. To measure the agreement and have a parameter for the comparison, a normalization to one was performed and the reduced $\chi^2$ was calculated. At first, simulations were performed considering the nominal momentum spread of 4 %. However, the first outputs showed a large deviation from the measured profile, especially in ARBY. In this case, two could be the source of error: an incorrect source-to-sample distance or the momentum spread value. Considering that for a fixed source-to-sample distance the simulated profile was narrower than the measured one, the only parameter that could influence the simulation was the momentum spread. Therefore, this value was increased to 5 %. With a higher momentum spread the simulations provided better results, with an improved agreement between measurements. The variation of the reduced chi-square value as a function of momentum is reported in Table 1 of the supplementary material section. From this table, it is also possible to have an error estimation of the thickness of the gold layer. The reduced chi-square value is almost equal for differences of 0.5 μm, while varies significantly for a 1 μm variation. It follows that the gold layer size can be addressed with an uncertainty of ± 1 μm. For sample A, the best fit was reached with a simulated gold layer of 3.5 ± 1 μm, in agreement with SEM results (Fig. 4a). Here, ARBY resulted in a better fit than TRIM, especially for the higher momentum runs, where TRIM results in a broader profile. The broader profile of the TRIM simulation is also evident in sample B, where the agreement with the measured data is worse than ARBY (Fig. 4b). Here, the best fit was reached at a gold thickness of 4.5 ± 1 μm. Sample C, instead, with an SEM average thickness of 7.3 ± 0.8 μm is well-fitted by both software (Fig 4c). In this case, along with the momentum spread, some adjustments had to be made in terms of source-to-sample distance. In particular, the best fit was reached at a closer distance of 9.7 cm. This was confirmed by comparing the aluminium profile of the three measured foils: for sample C, a small difference in the measured profile suggested a change in distance. As shown in Figure 1, sample positioning is done by hanging the samples onto an aluminium holder. Despite the best efforts to place each standard in the same position, small errors could have occurred, thus influencing the simulation output. The adjustment provided a better result, and the gold layer was identified to be 7.5 ± 1 μm, in agreement with the SEM data. Finally, even if the focus of the analysis is the gold layer, it is worth showing an example of the other layers, in particular nickel. Figure 4d reports the nickel layer modelling of the sample. In electroplating, nickel is mostly used to prevent degradation: this element is more durable and resistant to oxidation than copper and acts as a protective layer. As the nickel is deposited quite quickly (with a "nickel flash") this layer is not always even, but from SEM scans it was set to have an average thickness of about 8 μm. As shown in the insight of Figure 4d, while EP_A and EP_B nickel profiles are almost the same, for EP_C the profile is shifted to higher momentums. This is because, in EP_C, muons travel a significantly thicker layer of gold before entering the nickel layer. In the supplementary material section, the modelling plots of all the other layers of the samples are reported. It is worth mentioning that samples A and B have a quite thin layer of gold, close to the minimum thickness that the software (around 2 μm) is able to model with this set of momentum values. Due to the size of this layer, the modelling of these samples, especially in TRIM, was quite challenging and the results have discrepancies from the measured profile. However, this extreme condition of the modelling process sets the lower limit that can be reached with the simulations, as the results improve with the increase of the gold layer.

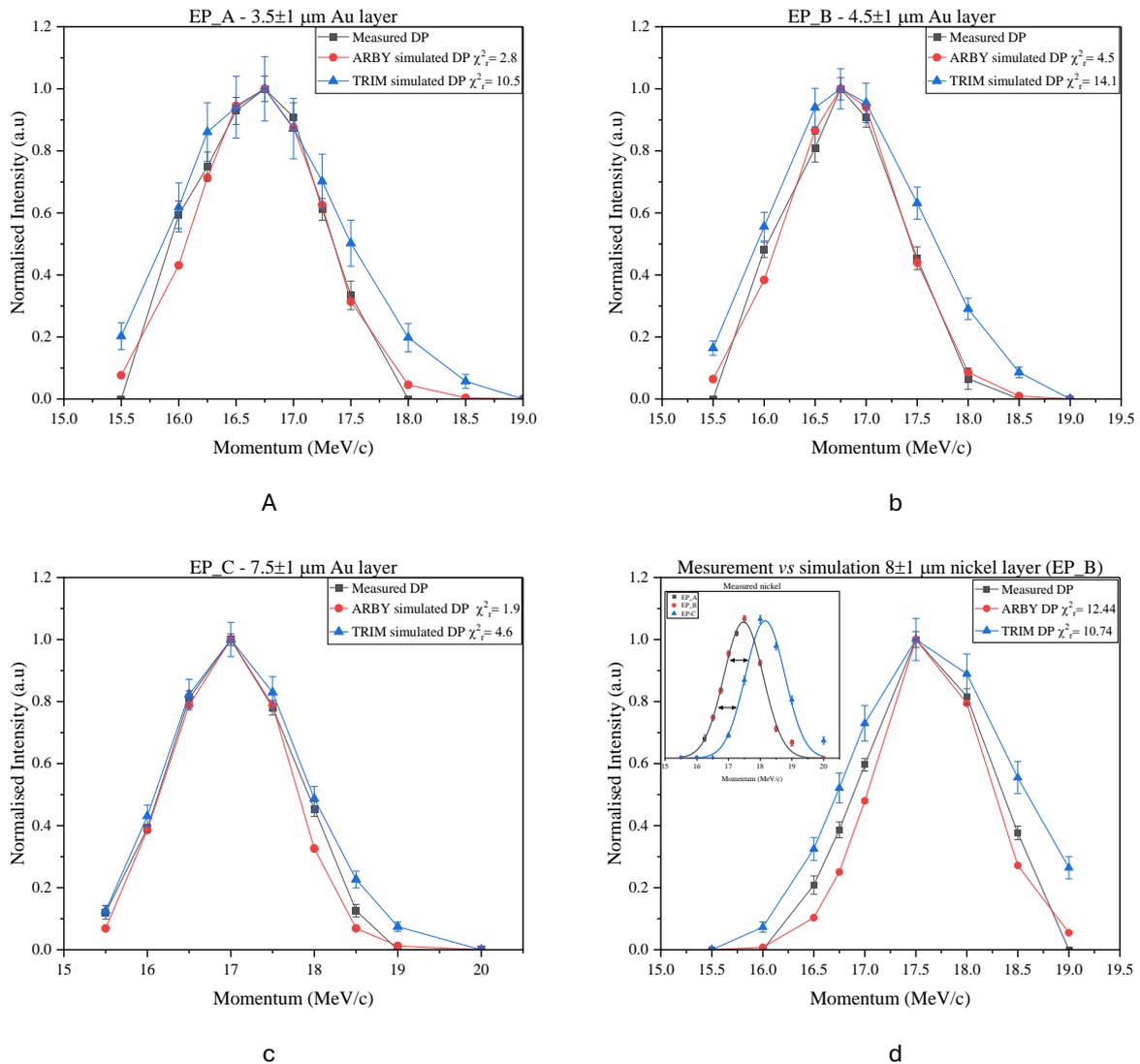

***Figure 4*** *– Electroplated samples, evaluation of the gold layer thickness. a) sample EP_A; b) sample EP_B; c) sample EP_C. For a thin layer of gold, TRIM performances are worse than ARBY, with a larger deviation from measured data. The results align in sample C, where the gold layer is thicker. D) simulation of the nickel layer with ARBY and TRIM. In the insight graph, the measured profile of nickel in the three samples: for EP_C (blue) the profile is shifted to a higher moment due to the increased size of the gold layer.*

### 2.2 Amalgam samples

Differently from the electroplated samples, the amalgam gildings are handmade. Made in collaboration with the *Opificio delle Pietre Dure,* this set of samples was made following the medieval recipe of Benvenuto Cellini. Here, the gold was applied to bronze (EM2) and brass (SM3). Due to the complex process of amalgam gilding, it was difficult for the craftsmen to have complete control of the procedure. As shown in Table 1, the gold film is characterised by the presence of cracking and air pockets, mostly due to the evaporation of mercury. This effect is generally mitigated by polishing and burnishing, but in this case, some defects were still visible. It results that, differently from the previous foils, the two samples have a different gold depth profile (Figure 5). Especially for sample EM2 whose profile is distant from the well-defined gold profile reported in Figure 3. For this sample, when performing the preliminary set of simulations with a gold layer size of 10 μm, the simulated outputs had a large deviation from the real measurement, suggesting a thinner layer of gold. Therefore, to take into account a reduction of the gold layer, simulations were performed with a decreased thickness. While this improved the results of the simulations, the agreement with the measurement was quite poor. Hence, to take into account the presence of air bubbles in the layer and replicate them in the simulations, the standard density of the gold (19.32 g/cm³) was reduced by a defined percentage (from 5 % to 20 %). Especially with TRIM, where there is no room for modelling, this stratagem helped with the simulation and provided results that fit better the experimental data.

For EM2 the best fit was reached with a gold thickness of 5 ± 1 μm and the density of the material decreased by 20 % from the nominal value of gold (a reduced $\chi^2$ of 2.8 for SRIM-TRIM and 2.9 for ARBY was obtained, as shown in Figure 5a). For sample SM3, instead, size and density parameters were not changed as the preliminary simulations provided results with reasonable agreement with the measurement. Here, only size was changed, in a range from 8 to 12 μm. The best fit was reached with a thickness of 11 ± 1 μm and standard gold density, in agreement with the SEM measurement, as shown in Figure 5b. As for the electroplated samples, the plots of the other layers are reported in the supplementary material section.

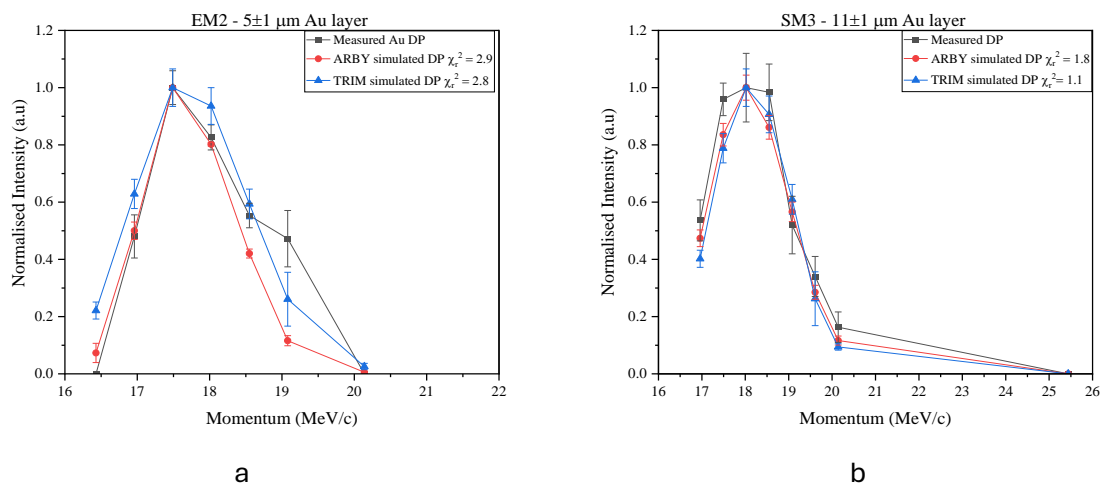

*Figure 5 – Amalgam samples. a) for the EM2, a reduction in thickness and density of the layer was required to reach a reasonable agreement with the experimental data; b) for the SM3 sample, the simulations were performed by only varying the size of the gold layer and the best fit was reached at 11 μm are, in agreement with SEM measurement.*

## Discussion

This work wanted to prove the goodness of μ-XES depth profiling for Heritage Science artefacts: the measurement provided a full characterisation of all the layers present in the sample, with a particular interest in gold. Moreover, the analysis and interpretation of the collected data with the use of simulation software provided promising results. For both sets of samples, ARBY and TRIM were able to replicate the experimental results, as reported in Table 2. Differences in the results are defined by the reduced chi-square value, which measures the agreement between modelling and measurement. This parameter can be addressed in numerous ways: here, the calculations take into account also the momentum runs where the measured output is zero (i.e. gold was not detected). For this reason, the best values of chi-square are around three for the EP and EM2 samples (that have runs where gold was not detected), while closer to one for the SM3 sample. The results of the EP_A and EP_B samples suggest that ARBY is capable of modelling finer steps in thickness and thinner layers compared to TRIM. For these two samples, the TRIM simulated profile has a broader tail which causes the discrepancy with the measured data. An increased effort was required for sample EM2, for which a large number of simulations were performed before finding the best fit. In this case, the unevenness of the surface made the modelling process more complicated. The results, however, suggest that by making some assumptions (i.e. reducing thickness or modifying density), the two software can handle uneven surfaces. With a thicker and even layer of gold, instead, the results of the two software align: the EP_C and SM3 samples are well-fitted by both software. In addition, for each of the electroplated samples, it was possible to model the nickel layer with reasonable agreement with the measured data. Overall, the results suggest that TRIM performances with thin layers are worse than ARBY, but they improve with an increase in thickness. For both software, the difference in the simulated output can be ascribed to two main causes (apart from gold size considerations): the momentum spread and the sample position. For the first, which is also linked to how the beam profile is implemented in the software, a variation of the nominal value is plausible (from 4 to 5 %), especially after the facility's recent refurbishment. This aspect was not extensively investigated for this research as it would have required additional investigation with measurements at the facility, but the results are consistent with an increased value of spread. Therefore, with a small increase in spread, better results were found with the electroplated samples (which, in the time scale, were measured after the amalgam samples and after the refurbishment). The second source of discrepancy, sample positioning, is mostly linked to the experimental setup. As shown in Figure 1, the current sample positioning is always open to possible errors, that can influence the results. In sample C, for instance, it has been addressed that the beam window sample exit distance could have influenced the simulations. Therefore, especially with layered materials it is important to carefully place the sample in front of the beam exit and perpendicular to the beam direction. The setup has been used for many experiments, but for this specific case, where the modelling in simulation software was an important part of the experiment, it represented a source of error. These topics were not extensively investigated in this proof of concept but will be addressed in follow-up experiments at the facility. To conclude, the consistency of the results of the ARBY and TRIM software testifies to the quality of the approach, which with more improvements could provide even more reliable results. For instance, efforts have been made by developers to insert CAD files in the simulations, so complex geometries could be implemented. Currently, ARBY is available only for INFN Bicocca staff, but discussions are ongoing to make this software available to the general users. Anyway, ARBY is GEANT4, so users can create their own applications to run the same simulations. TRIM, instead, does not have all the modelling tools as ARBY, but it is open-source and easy to handle, making it ready to use. What emerges from this characterisation, in the end, is that besides diversities in the simulation process, both software can provide a reliable source for better data interpretation.

**Table 2** – Comparison between measurements and simulations (ARBY and TRIM)

| Sample | Measured thickness (SEM - µm) | Calculated thickness (Simulation - µm) |
|---|---|---|
| EPA | 3.3 ± 0.2 | 3.5 ± 1 |
| EPB | 4.6 ± 0.6 | 4.5 ± 1 |
| EPC | 7.3 ± 0.8 | 7.5 ± 1 |
| SM3 | 11 ± 1 | 11 ± 1 |
| EM2 | | 5 ± 1 |

## Conclusions

Muonic X-ray Emission spectroscopy is a powerful method for the non-invasive characterisation of Heritage science artefacts. In this study, a protocol for the data analysis of depth profile experiments with negative muons was proposed. By coupling the muonic X-ray spectra analysis with the use of Monte Carlo simulation software, the work aimed at assessing the thickness of a superficial gold layer in mock-up samples. The two software used in this proof of concept, despite differences in the simulation process, provided satisfactory results, in agreement with the preliminary characterisation of the samples. For all the investigated samples, the size of the gold layer was addressed. ARBY, with the possibility of a more detailed modelling of the sample, provided better results than TRIM. TRIM, however, is open-source software that can provide quick and reliable results even to a non-expert user, while ARBY would require extensive training. The choice of one software over the other would be a trade-off between different aspects, mostly depending on the capabilities of the users. However, the work demonstrates that the results provided by the two software are consistent with each other. To minimize differences between simulations and measurements, it would be important to have a better understanding of the simulated beam shape as well as the beam spread, which could be responsible for the broadening of the simulated depth profile. Moreover, it would be interesting to find a lower limit of this method in terms of the minimum detectable thickness of a given layer, gold in this specific case. Finally, in terms of experimental procedure, it would be important to reduce some of the sources of errors, like the positioning of the samples in the beam. To address these issues, more tests of mock-up samples are foreseen in the future, with planned beam time at the ISIS Neutron and Muon source.

## Funding

## Acknowledgements

Experiments at the ISIS Neutron and Muon Source were supported by a beamtime allocation RB1910123 and RB2300032 from the Science and Technology Facilities Council. Data is available here: https://doi.org/10.5286/ISIS.E.RB1910123 and https://doi.org/10.5286/ISIS.E.RB2300032-1. The LEA laboratory of the University of Firenze and Dr. Arianna Meoli are also acknowledged for the manufacturing of the electroplated sample and the SEM analysis.

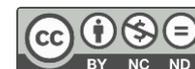

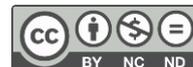